# Integrating Artificial Intelligence with Real-time Intracranial EEG Monitoring to Automate Interictal Identification of Seizure Onset Zones in Focal Epilepsy

**Authors**

Yogatheesan Varatharajah[1], Brent Berry[2,5], Jan Cimbalnik[3], Vaclav Kremen[2,3], Jamie Van Gompel[4], Matt Stead[2], Benjamin Brinkmann[2,5], Ravishankar Iyer[1], and Gregory Worrell[2]

[1] Electrical and Computer Engineering, University of Illinois, Urbana, IL 61801, USA.
[2] Mayo Systems Electrophysiology Laboratory, Department of Neurology, Mayo Clinic, Rochester, MN 55905, USA
[3] Czech Institute of Informatics, Robotics and Cybernetics, Czech Technical University in Prague, 166 36 Prague 6, Czech Republic
[4] Department of Neurosurgery, Mayo Clinic, Rochester MN, 55905, USA
[5] Department of Physiology & Biomedical Engineering, Mayo Clinic, Rochester MN, 55905, USA

**Abstract**

An ability to map seizure-generating brain tissue, i.e., the *seizure onset zone* (SOZ), without recording actual seizures could reduce the duration of invasive EEG monitoring for patients with drug-resistant epilepsy. A widely-adopted practice in the literature is to compare the incidence (events/time) of putative pathological electrophysiological biomarkers associated with epileptic brain tissue with the SOZ determined from spontaneous seizures recorded with intracranial EEG, primarily using a single biomarker. Clinical translation of the previous efforts suffers from their inability to generalize across multiple patients because of (a) the inter-patient variability and (b) the temporal variability in the epileptogenic activity. Here, we report an artificial intelligence-based approach for combining multiple interictal electrophysiological biomarkers and their temporal characteristics as a way of accounting for the above barriers and show that it can reliably identify seizure onset zones in a study cohort of 82 patients who underwent evaluation for drug-resistant epilepsy. Our investigation provides evidence that utilizing the complementary information provided by multiple electrophysiological biomarkers and their temporal characteristics can significantly improve the localization potential compared to previously published single-biomarker incidence-based approaches, resulting in an average area under ROC curve (AUC) value of 0.73 in a cohort of 82 patients. Our results also suggest that recording durations between ninety minutes and two hours are sufficient to localize SOZs with accuracies that may prove clinically relevant. The successful validation of our approach on a large cohort of 82 patients warrants future investigation on the feasibility of utilizing intra-operative EEG monitoring and artificial intelligence to localize epileptogenic brain tissue. Broadly, our study demonstrates the use of artificial intelligence coupled with careful feature engineering in augmenting clinical decision making.

**Introduction**

Epilepsy is one of the most prevalent and disabling neurologic diseases. It is characterized by the occurrence of unprovoked seizures and affects ~1% of the world's population [Leonardi 2002]. Many patients with epilepsy achieve seizure control with medication, but approximately one-third of people with epilepsy continue to have seizures despite taking medications [Kwan 2000]. In such cases, one treatment option is surgical resection of the brain tissue responsible for seizures, but this option depends critically on accurate localization of the pathological brain tissue, which is referred to as the *seizure onset zone* (SOZ). Clinical SOZ localization requires implanting of electrodes for intracranial EEG (iEEG) that is recorded over several days to allow sufficient time for spontaneous seizures to occur [Lüders 2006]. The electrodes that are in the SOZ are identified based on visual inspection of the iEEG captured at the time of seizures, and some tissue around these electrodes is removed during a surgical procedure. Despite being the current gold standard for mapping of the epileptic brain in a clinical setting, this manual procedure is time-consuming, costly, and associated with potential morbidity [Van Gompel 2008, Wellmer 2012]. Recently, the use of interictal (non-seizure) iEEG data for the identification of SOZs and for the possibility of replacing multi-day ICU monitoring to record habitual seizures has received notable interest [van 't Klooster 2015].

    A common practice undertaken in the literature investigating interictal SOZ localization is to compare the incidence rates (events/time) of putative pathological electrophysiological events associated with epileptic brain tissue (known as *electrophysiological biomarkers* of epilepsy) detected in iEEG recorded from individual electrodes against the gold-standard SOZ electrodes determined from spontaneous seizures. Among the potential electrophysiological biomarkers, high-frequency oscillations (HFOs) [Bragin 1999, Bragin 2002, Worrell 2011, Zijlmans 2012] and interictal epileptiform discharges (IEDs) [Staley 2011] have been the most widely investigated. Phase-amplitude coupling (PAC) and other forms of cross-frequency coupling (CFC) have more recently been investigated as promising clinical biomarkers for epilepsy [Papdelis 2016]. HFOs are local field potentials that reflect short-term synchronization of neuronal activity, and they are widely believed to be clinically useful for localization of epileptic brain [Bragin 1999, Worrell 2004, Jirsch 2006, Staba 2002]. Furthermore, there is an extensive literature investigating IEDs as interictal markers of seizure onset zones, but it has met with limited success [Lüders 2006, Marsh 2010, Hauf 2012]. PAC (a measure of cross-frequency coupling) [Jensen 2007] as an adjunct to ictal (seizure) biomarkers was shown to be useful for SOZ localization [Edakawa 2016], and more recently, PAC has been evaluated as an interictal marker for determining SOZs [Amiri 2016]. Most the existing studies have utilized a simple counting of the above biomarkers (detected either manually or using software) in fixed durations to classify the electrodes that are in the SOZ [Cimbalnik 2017]. Although some recent approaches have utilized clustering methods [Liu 2016] and dimensionality reduction methods [Weiss 2016] as preprocessing steps in identifying pathologic interictal HFOs, the determination of SOZs was still performed using simple counting of HFOs. Furthermore, these approaches have predominantly utilized a single biomarker to identify SOZs and have not considered the inter-patient variability nor the temporal dynamics of the epileptic activity [Cimbalnik 2017]. As a result, they have not been able to generalize across multiple patients and their overall accuracies have been insufficient to bring them into clinical practice [Holler 2015, Nonoda 2016, Sinha 2017].

    The reasons for inter-patient variability might include electrode placement, false-positive detections of biomarkers, signal artifacts, the varied etiology of focal epilepsy, or the fact that some biomarkers are incident in both physiologic and pathologic states [Matsumoto 2013, Ben-Ari 2007, Cimbalnik 2016]. Thus, utilizing a single biomarker to identify SOZs of patients with potentially heterogeneous epileptogenic mechanisms may result in unsatisfactory accuracy for some individuals. We hypothesize that it may be possible to reduce inter-patient variability by combining the complementary values contained within different electrophysiological biomarkers and thereby improve SOZ localization potential. However, despite the growing interest in each of the above biomarkers, the extent to which they provide independent predictive value for epileptogenic tissue localization remains unclear. From a signal-processing perspective, IED represents a relatively distinct electrophysiological phenomenon compared to HFO and PAC. However, temporal correlations of HFO events and PAC may be observed when short HFOs co-occur with IEDs [Weiss 2016]. Apart from that specific instance, it is possible that each biomarker will constitute specific

electrophysiological information about the epileptogenicity of brain tissue and might add predictive value when used in unison with other biomarkers. However, the potential clinical utility of combining electrophysiological biomarkers has received relatively little investigation [Gnatkovsky 2014].

In addition, there is evidence that behavioral states play a role in altering the temporal patterns of epileptiform activity in the brain [Staba 2002, Worrell 2008, Amiri 2016]. As a result, the occurrence of the biomarkers exhibits temporally varying rates when long EEG recordings with mixed behavioral states are considered [Pearce 2013]. Hence, the common practice of simply counting HFOs or IEDs for a fixed duration and using an average rate to determine the SOZ is likely suboptimal. We recently proposed a temporal filtering-based unsupervised approach to utilize temporal characteristics of spectral power features to determine SOZs interictally [Varatharajah 2017]. However, more sophisticated models are needed to effectively utilize multiple electrophysiological biomarkers and their temporal characteristics to accurately determine SOZs. Modern artificial intelligence (AI) based methods facilitate (a) the ability to learn high-dimensional decision functions from labeled training data and (b) the flexibility to define customized features representing domain knowledge [Russell 1995]. We believe that these properties can be useful in harnessing multiple electrophysiological biomarkers and their temporal characteristics for the interictal classification of SOZs. However, AI-based approaches have been underexplored in the SOZ classification literature, mainly because of the unavailability of large-scale iEEG datasets collected during epilepsy-surgery evaluation. Fortunately, the availability of continuous iEEG recordings collected from a large cohort of 82 patients and approximately 5000 electrodes gives us a unique opportunity to assess the potential utility of AI-based approaches in this study. Although this dataset is still not large enough to automatically extract class-specific electrophysiological patterns using deep-learning approaches, comparable performance can be realized on this dataset using careful feature engineering and appropriate model selection.

In that context, the aim of this study is to develop an AI-based analytic framework that utilizes multiple interictal electrophysiological biomarkers (e.g., HFO, IED, and PAC) and their temporal characteristics for interictal electrode classification and mapping of SOZs. To that end, we developed a support vector machine (SVM) based classification model utilizing customized features (based on the above biomarkers) extracted from 120-minute interictal iEEG recordings of 82 patients with drug-resistant epilepsy to interictally identify the electrodes representing their seizure onset zones. This approach achieved an average AUC (area under ROC curve) of 0.73 when the HFO, IED, and PAC biomarkers were used jointly, which is 14% better than the AUC achieved by a conventional biomarker incidence-based approach using all three biomarkers, and 4–13% better than that of an SVM-based model that used any one of the biomarkers. This result indicates that exploiting the temporal variations in biomarker activity can improve localization of epileptic brain and that the biomarkers utilized in this study have complementary predictive values. Our analysis of individual patients reveals that the AUCs improve or remain unchanged for more than 65% of the patients when the composite, rather than any single biomarker, is utilized, supporting the hypothesis that combining multiple biomarkers can provide more generalizability than can individual biomarkers. Development of this technology also enabled us to develop an understanding of the recording durations required for interictal localization of SOZs. By analyzing iEEG segments of different durations (10–120 minutes), we show that longer iEEG segments provide better accuracy than do short segments, and that the improvements become statistically insignificant for durations beyond 90 minutes. These promising results warrant further investigation on the feasibility of using intra-operative mapping to localize epileptic brain.

**Methods**

*Experimental Setup*

Data used in this study were recorded from patients undergoing evaluation for epilepsy surgery at the Mayo Clinic, Rochester, MN. The Mayo Clinic Institutional Review Board approved this study, and all subjects provided informed consent. Subjects underwent intracranial depth electrode implantation as part of their evaluation for epilepsy surgery whenever noninvasive studies could not adequately localize the SOZ. To provide an unbiased dataset for analysis, we took 2 hours of continuous iEEG data (during some period between 12:00 and 3:00 AM) on the night after surgery.

*Subjects*

Data from 82 subjects (48 males and 34 females, with an average age of 31) with focal epilepsy were investigated by post hoc analysis. All subjects were implanted with intracranial depth arrays, grids, and/or strips; see supplementary Table 1 for details. Subjects underwent multiple days of iEEG and video monitoring to record their habitual seizures.

*Electrodes and Anatomical Localization*

Depth electrode arrays (from AD-Tech Medical Inc., Racine, WI) were 4- and 8-contact electrode arrays consisting of a 1.3-mm-diameter polyurethane shaft with platinum/iridium (Pt/Ir) macroelectrode contacts. Each contact was 2.3 mm long, with 10-mm or 5-mm center-to-center spacing (with a surface area of 9.4 mm$^2$ and an impedance of 200–500 Ohms). Grid and strip electrodes had 2.5-mm-diameter exposed surfaces and 1-cm center-to-center spacing of adjacent contacts. Anatomical localization of electrodes was achieved using post-implant CT data co-registered to the patient's MRI using normalized mutual information [Ashburner 2008]. Electrode coordinates were then automatically labeled by the SPM Anatomy toolbox, with an estimated accuracy of 0.5 mm [Tzourio-Mazoyer, et al., 2002].

*Signal Recordings*

All iEEG data were acquired with a common reference using a Neuralynx Cheetah electrophysiology system. (It had a 9-kHz antialiasing analog filter, and was digitized at a 32-kHz sampling rate, filtered by a low-pass, zero-phase-shift, 1-kHz, low-pass Bartlett-Hanning window, and down-sampled to 5 kHz.)

*Clinical SOZ Localization*

The SOZ electrodes and time of seizure onset were determined by visually identifying the electrodes with the earliest iEEG seizure discharges. Seizure onset times and zones were determined by visual identification of a clear electrographic seizure discharge, followed by looking back at earlier iEEG recordings for the earliest electroencephalographic change contiguously associated with the visually definitive seizure discharge. The same approach has been used previously to identify neocortical SOZs [Worrell 2004] and medial temporal lobe seizures [Worrell 2008]. The identified SOZ electrodes were used as the gold standard to test and validate our analyses.

*Data*

Continuous 2-hour interictal segments of iEEG data, sufficiently separated from seizures, were chosen for all 82 patients to represent a monitoring duration that could be achieved during surgery. A total of 4966 electrodes were implanted across the 82 subjects, and 911 of them were identified to be in SOZs via ictal localization performed by clinical epileptologists caring for the patients.

*Data Preprocessing*

Prior to analysis, continuous scalp and intracranial EEG recordings were reviewed using a custom MATLAB viewer [Brinkmann 2009]. Electrode channels and time segments containing significant artifacts or seizures were not included in subsequent analyses. All iEEG recordings were filtered to remove 60-Hz power-line artifacts.

*Overall Analytic Scheme*

Selected 2-hour iEEG recordings were divided into non-overlapping 3-second epochs. A 3-second epoch length was chosen to accommodate at least a single transient electrophysiologic event (in the form of a PAC, HFO, or IED) that could be associated with the SOZ. The HFO [Cimbalnik 2017], IED [Barkmeier 2012], and PAC [Amiri 2016] biomarkers were extracted using previously published detectors to measure their presence in each 3-second epoch. Then, a clustering procedure was performed to assign a binary observation of *normal* or *abnormal* to each channel. This procedure was performed separately for each patient, and the biomarker measures extracted in a 3-second recording of all the channels of a specific patient were considered. These channels were clustered into two groups based on their similarities with respect to each biomarker, and the cluster with the larger average biomarker rate was considered the *abnormal* cluster. This step was performed so that biomarker detections that had strong magnitudes and showed strong spatial correlation were retained, and electrodes with noisy detections were minimized. At the end of that procedure, every 3-second recording of a channel was associated with three binary values (one for each biomarker) representing the presence of the HFO, IED, and PAC biomarkers. We refer to those binary values as *observations*. Since there are 2400 3-second epochs in a 2-hour period, the total number of observations made in a channel was 3 * 2400 = 7200. Since that number of features is relatively large compared to the number of channels available in our study, we reduced the number of observations by applying a straightforward summation approach. Binary observations made within a 10-minute window (200 epochs) were counted to arrive at a measure of the local rate of the biomarker incidence for each 10-minute window of the 2-hour recording. Although the 10-minute window length may appear to have been chosen arbitrarily, it is large enough to have less noisy local biomarker incidence rates and yet not so large as to mask the temporal variations in interictal biomarker activity. This method reduces the number of observations for a channel to 36 (3 local biomarker rates × 12 windows). These observations, made across a 2-hour period of a channel, were used to infer whether that channel belonged to an SOZ under a supervised learning setting using a support vector machine (SVM) classifier. The whole process is illustrated as a flow diagram in Figure 1.

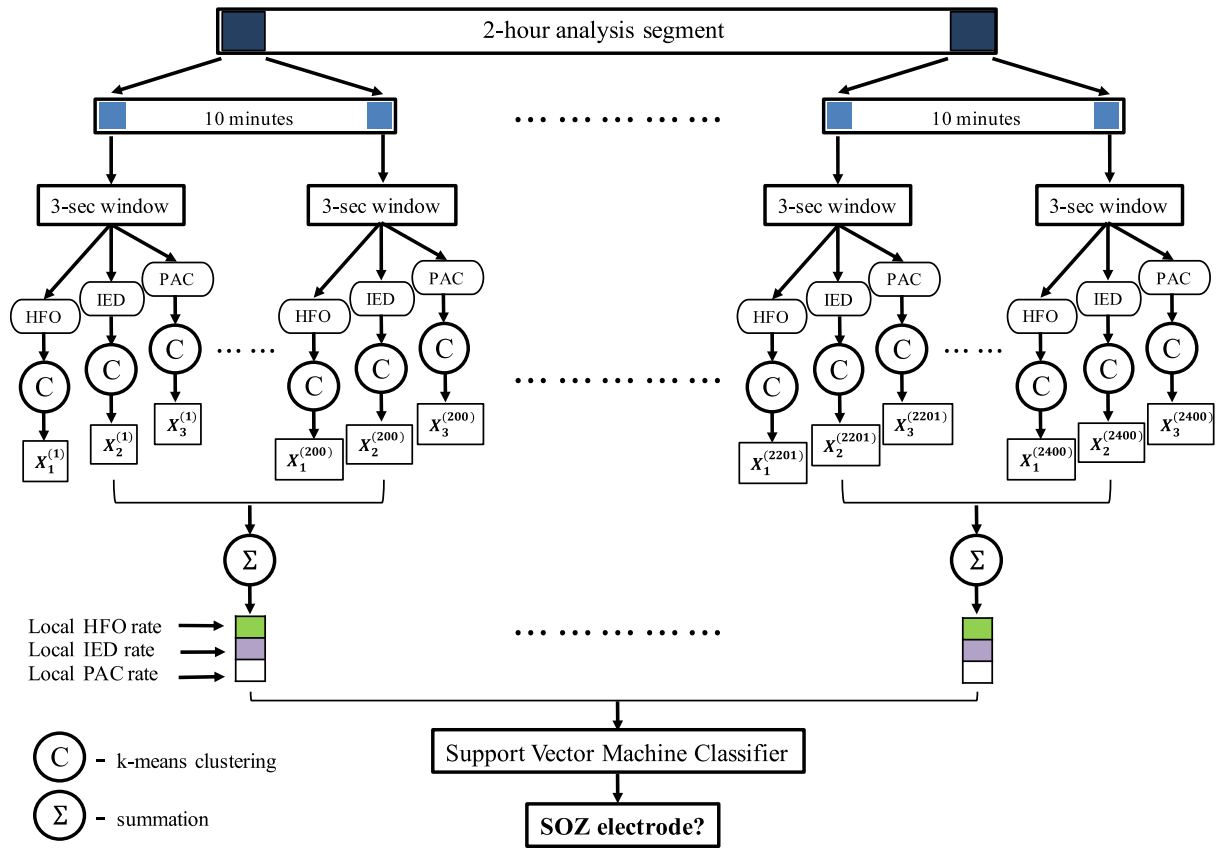

**Figure 1: The overall analytic scheme of the SOZ detection algorithm utilized in this study.** A 2-hour data segment is analyzed for each patient. PAC, HFO, and IED biomarkers are extracted in 3-second epochs, and a clustering method is used to group channels based on similarities with respect to the biomarkers. These groupings are converted to binary (0, 1) observations and counted within a 10-minute window to obtain local biomarker incidence rates. These local biomarker incidence rates for the three biomarkers within all the 10-minute windows of a channel's 2-hour recording are utilized as the features of that channel in a machine learning setting. A support vector machine (SVM) classifier, which was trained and tested using labeled training data, is used to predict whether an electrode is in an SOZ.

*Detection of Interictal Electrophysiological Biomarkers*

The PAC measure was calculated by correlating instantaneous phase of the low-frequency signal with the corresponding amplitude of a high-frequency signal for a given set of low- and high-frequency bands (Figure 2a). In this implementation, low- and high-frequency contents in the signal were extracted using MORLET wavelet filters, and all frequency bands were correlated against all others to create a so-called *PAC-gram* (Figure 2b). Based on the observed high PAC content and the existing literature [Weiss 2015], 0.1–30 Hz was chosen as the low-frequency (modulating) signal, and 65–115 Hz was chosen as the high-frequency (modulated) signal in the rest of the analysis. HFOs were detected using a Hilbert transform-based method [Kucewicz & Berry 2015, Pail 2017, Cimbalnik 2016]. The data segments were bandpass-filtered for every 1-Hz band step from 50 to 500 Hz. Then, the filtered-data frequency bands were normalized (z-score), and the segments in which the signal amplitudes were three standard deviations above the mean for a duration of one complete cycle of a respective high frequency (in 65–500 Hz) were marked as HFOs (see Figure 2c) [Matsumoto 2013]. IEDs were extracted using a previously validated spike-detection algorithm [Barkmeier 2012]. A detection threshold of four standard deviations (of differential amplitude) around the mean was used to mark IEDs in this algorithm (see Figure 2d). The HFO, IED, and PAC detected events were stored in a database.

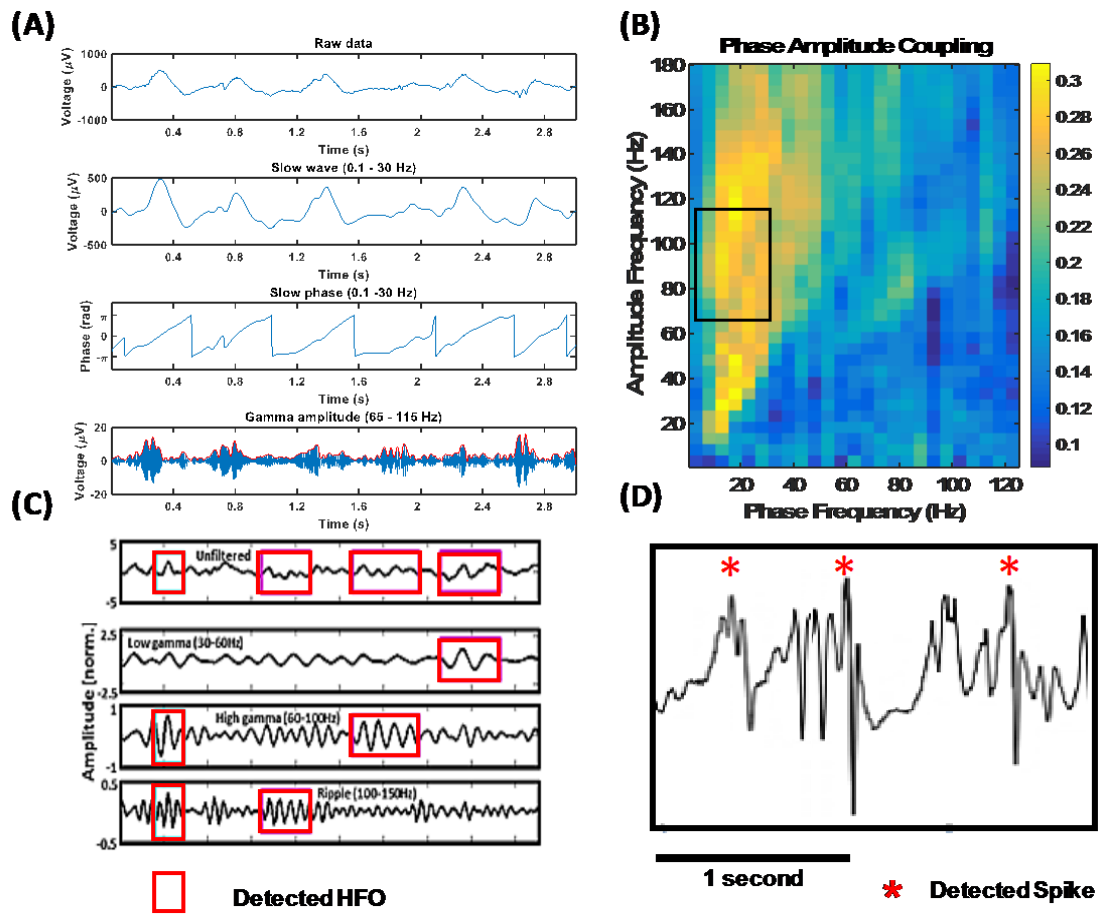

**Figure 2: Phase amplitude coupling (PAC), high-frequency oscillations (HFO), and interictal epileptiform discharge (IED) detection.** **(A)** Detailed illustration of the PAC feature extraction algorithm. Low (0.1–30 Hz) and high (65–115 Hz) frequency components are filtered out from the raw signal. The phase of the slow wave is correlated with the high-frequency amplitude envelope to measure coupling. **(B)** A PAC-gram representing the average interictal PAC measured between different frequency bands. Highlighted portion indicates the low- and high-frequency bands utilized in the rest of our analysis. **(C)** Pictorial illustration of HFO detection. Oscillations that have an amplitude of three standard deviations above the mean and lasting for more than one complete cycle in low-gamma (30–60 Hz), high-gamma (60–100 Hz), and ripple (100–150 Hz) bands are detected. **(D)** An illustration of detected IEDs. Differential amplitude is standardized, and a threshold of four standard deviations around the mean was used to mark IEDs.

*Prediction of SOZ Electrodes Using a Support Vector Machine Classifier*
The biomarkers extracted from a 2-hour recording of a channel were converted to a 36-dimensional feature vector as shown in Figure 1. The features represent the local biomarker incidence rates (with a separate rate for each biomarker) within each 10-minute window of the 2-hour recording. These features were standardized to eliminate any differences in scale. We took two different approaches to perform cross-validation. First, we performed 10-fold cross validation. The dataset, including standardized features of 4966 electrodes (including 911 SOZ electrodes), was divided into a 60% training set and a 40% testing set, keeping the same proportion of SOZ and NSOZ (non-SOZ) electrodes in both sets. Second, we performed leave-1-out cross-validation. For every subject in the dataset, we used the data from the rest of the subjects as the training data and the respective subject's data as the testing data. In each of the cross-validation iteration, training and testing datasets were generated using one of the cross-validation approaches. An SVM classifier was trained on the training set, whose hyper-parameters (described below) we optimized by performing a grid search with a tenfold cross-validation within the training set. The classifier trained on the best-performing hyper-parameters was used to predict the labels of the channels in the testing set. By comparing those predictions against the ground-truth labels of the testing set channels, we calculated the metrics of model fitness. This process was repeated 10 times with different combinations of training and testing sets to obtain metrics of generalized performance. This process is illustrated in Figure 3.

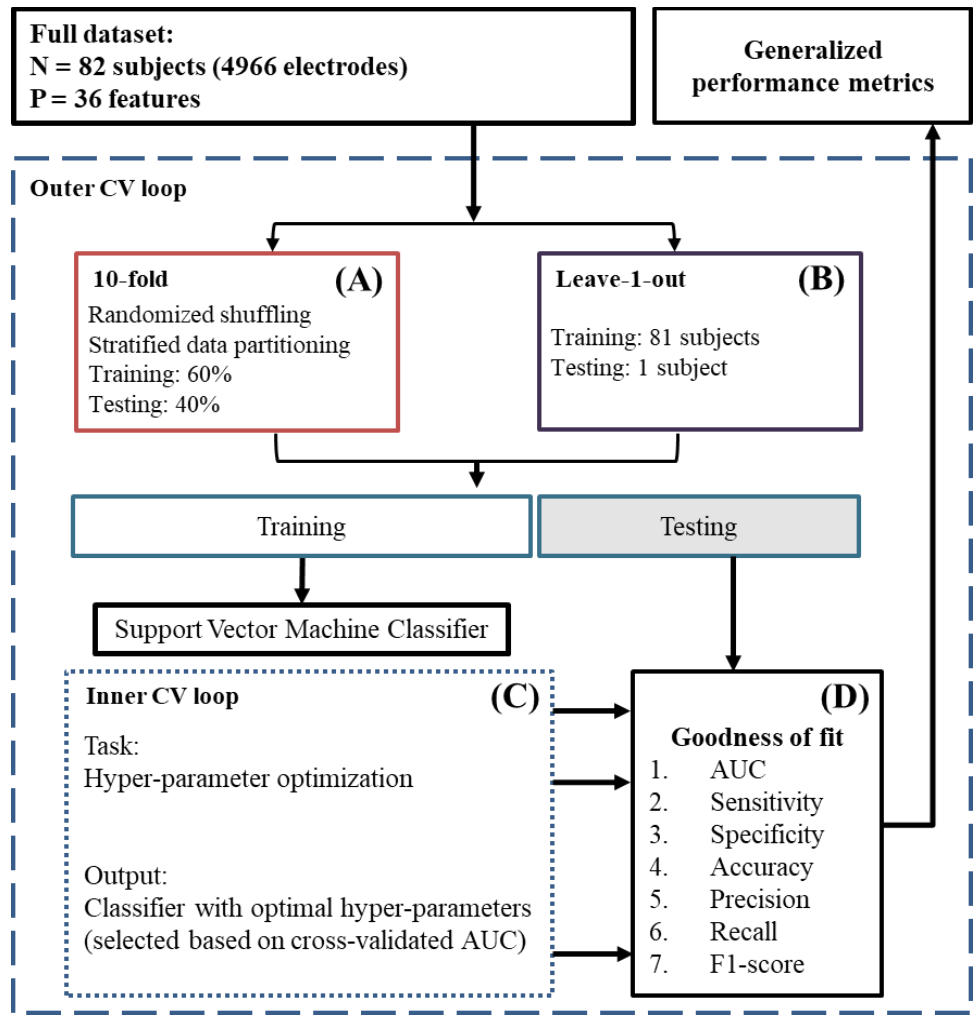

**Figure 3: A flow diagram illustrating the prediction framework.** The input is the whole dataset, including the 36 features extracted from 82 subjects (4966 channels) and their gold-standard labels assigned by clinical epileptologists. (A) 10-fold cross-validation: first, the dataset is shuffled to randomize the order of channels in it. After randomization, the dataset is partitioned into two sets, a 60% training set and a 40% testing set, keeping the same proportion of NSOZ and SOZ channels in both sets. (C) Inner CV loop: A set of optimal hyper-parameters is selected for the SVM classifier based on a tenfold cross-validation within the training set. (D) Goodness-of-fit metrics: The classifier learned in the previous step is tested on the testing dataset, and measures of its performance are generated. This whole procedure is repeated 10 times (i.e., 10-fold CV) or 82 times (i.e., leave-1-out CV) to produce generalized performance metrics, eliminating any bias introduced by a specific split of training and testing sets.

A **support vector machine** is a binary classifier that finds the maximum margin hyper-plane that separates the two classes in the data [Boser 1992]. The data being classified are denoted by $X \in \mathbb{R}^{N \times P}$ ($N$ channels and $P$ features), and the data from channel $i$ are denoted by $X^{(i)} \in \mathbb{R}^P$. The class labels for all the channels are denoted by $Y \in \{-1,1\}^N$ (where $-1$ and $1$ are numerical labels for the two classes), and the class label for channel $i$ is denoted by $Y^{(i)} \in \{-1,1\}$. The optimization problem to find the optimal hyper-plane (described by weights $W \in \mathbb{R}^P$ and intercept term $b \in \mathbb{R}$) is shown in Eq. 1.

$$\min_{W,b} \frac{1}{2}\|W\|^2 \quad (1)$$

$$\text{subject to } Y^{(i)}(W^T X^{(i)} + b) \geq 1, i \in \{1, \dots, N\}$$

Once the optimal hyper-plane $[W_{opt}, b_{opt}]$ is found, the predicted class label for channel $i$ is obtained as the sign of $W_{opt}^T X^{(i)} + b_{opt}$. This formulation assumes that the data have a clear separation between the two classes. When that is not the case, slack variables and a tolerance parameter (box-constraint) can be introduced to obtain separating hyper-planes that tolerate small misclassification errors [Cortes 1995].

Dual formulation of SVM has received considerable interest because it enables use of different kernel transformations of the original feature space without altering the optimization task and because of its advantages in complexity when the data are high-dimensional [Boser 1992]. Specifically, this allows for the features to be transformed from the original feature space to a kernel space. With this transformation, the cases in which the original data are not linearly separable may be solved because transformation of the data to higher dimensions may introduce linear separation in the transformed domain. Linear, radial basis function (RBF), and polynomial kernels are widely used kernels in this context.

**Goodness of fit** of the SVM classifier is evaluated by predicting the classes of the test dataset by using the classifier that was trained on the training dataset and comparing the predictions against the true class labels of the test dataset. This comparison is performed using standard performance metrics such as receiver operating characteristics (ROC) curve analysis, area under ROC curve (AUC), sensitivity, specificity, accuracy, precision, recall, and F1-score. Because of heterogeneity in the data, choosing one partition of training and test datasets is not sufficient to credibly evaluate the performance of a classifier. A common practice to obviate the effect of heterogeneity in the data is to perform several iterations of training-testing cross-validation of the dataset. One run of this procedure is carried out by choosing a subset of the dataset as training data, and testing on the rest of the dataset. This approach allows the calculation of generalizable performance metrics for the analyzed classifier.

**Results**

*Nonlinear Classification Boundary between SOZ and NSOZ Electrodes*

Understanding the nature of the separation between the two classes in feature space is important to achieve the maximum classification performance in binary classification. If the separation is linear, a linear classifier should be sufficient (and preferable due to the Occam's razor principle) to achieve the maximum attainable classification performance. On the other hand, when the separation is nonlinear, linear classifiers perform poorly compared to nonlinear classifiers. However, when the feature space is high-dimensional, visualizing the boundary between classes can be difficult. An option is to use linear and nonlinear classifiers to classify the two classes and plot the histograms of likelihood probabilities predicted by the classifier to understand the degree of separation achieved by linear and nonlinear boundaries [Cherkassky 2010]. We performed this analysis for our dataset by using an SVM classifier with linear and RBF kernels. Furthermore, we used the 10-fold cross-validation approach to perform this analysis because the individual AUCs obtained using the leave-1-out cross-validation approach were highly variable across patients. We trained two SVM classifiers with linear and RBF kernels, respectively, using the framework shown in Figure 3 with 60% of all the electrodes and all the biomarkers. These classifiers were used to predict the class labels for the rest of the electrodes (40%). We then compared the distribution of the likelihood probabilities generated by the two SVM classifiers against the true class labels. Figures 4a and 4b show the histograms of the likelihood probabilities obtained using an SVM classifier with linear and RBF kernels, respectively, for SOZ and NSOZ electrodes in the testing set. The nonlinear boundary achieved through an SVM classifier with an RBF kernel clearly has a better separation between the two classes, as can also be seen in Figure 4c. To quantify this observation, we plotted ROC curves (which are shown in Figure 4d) for the predictions obtained using linear and RBF kernels. The linear-SVM classifier obtained an AUC value of 0.57, while the RBF-SVM classifier obtained an AUC of 0.79 when all the biomarkers were utilized. Hence, we conclude that the boundary between SOZ and NSOZ electrodes is nonlinear in the feature space in which the features are derived as described in Figure 1, and the rest of our analyses focus on the results obtained by the SVM classifier with an RBF kernel.

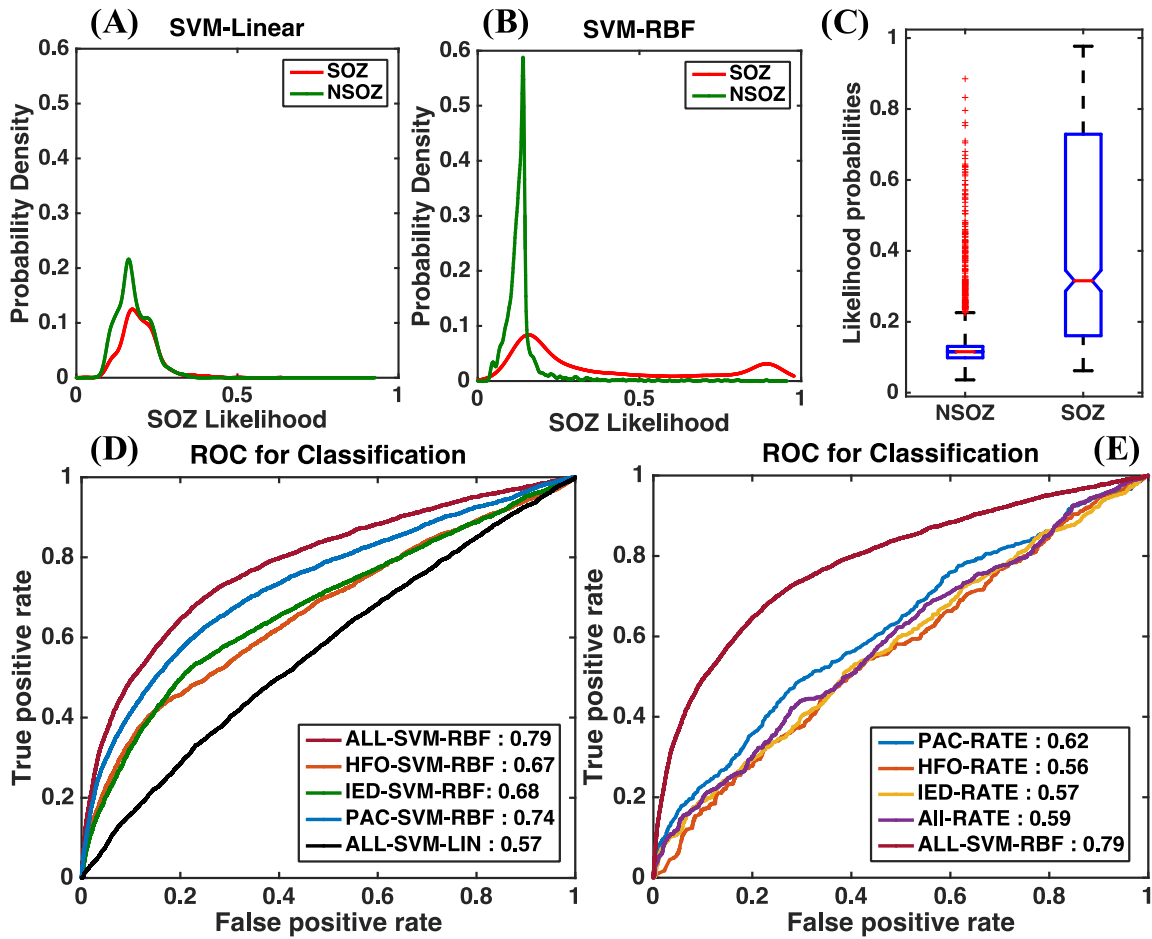

**Figure 4: Results obtained using a support vector machine with interictal electrophysiological biomarkers to classify seizure onset zone (SOZ) electrodes.** (A) and (B) show the probability densities of the likelihoods predicted by an SVM classifier for SOZ and NSOZ electrodes in the testing set for linear and RBF kernels, respectively. The RBF kernel results in less overlap between the SOZ and NSOZ probability densities. (C) Boxplot showing the range of likelihood probabilities obtained for SOZ and NSOZ electrodes when all biomarkers were used as features in an SVM classifier with an RBF kernel. (D) A comparison between the ROC curves when an SVM classifier was used with an RBF kernel for individual biomarkers and their combination, and when it was used with a linear kernel with a combination of all biomarkers. (E) A comparison between the AUCs obtained using conventional unsupervised methods that use overall rates of biomarker incidence to predict SOZ electrodes, and those obtained using our SVM-based supervised approach.

*A Supervised Learning Approach Improves the SOZ Localization Accuracy*

This paper proposes a supervised-learning-based approach that uses an SVM classifier to predict electrodes in an SOZ by using interictal iEEG data. In order to understand whether our supervised approach is better than simply using biomarker rates, we implemented biomarker-rate-based SOZ electrode classification for HFO, IED, and PAC separately and in combination. (We simply added the individual biomarker incidence rates to obtain an overall biomarker incidence rate.) Figure 4e shows a comparison between the ROC curves obtained for the unsupervised biomarker incidence rate-based approach and the supervised approach that uses an SVM classifier with an RBF kernel when all biomarkers were utilized. Predictions using the unsupervised approach were performed on the same testing set electrodes that were used in the supervised approach. The SVM-based supervised approach outperformed the unsupervised approaches with a 17–23% gain in the AUC value. Notably, the performance of the SVM classifier with a linear kernel was comparable to that of the unsupervised approach, with an AUC value of 0.57, as seen in Figure 4d. This highlights the ability to significantly improve the correct classification of previously unseen SOZ electrodes by utilizing the right machine learning method (in this case, SVM with an RBF kernel) to learn the characteristics of SOZ electrodes.

*Goodness-of-Fit Metrics for SOZ Electrode Classification*

Other goodness-of-fit metrics, as specified previously, were calculated for the different combinations of biomarkers and methods and are listed in Table 1. For all the metrics other than AUC, the likelihood probabilities assigned by the classifier were applied with a threshold to classify SOZ and NSOZ electrodes. In order to compare the different approaches, this threshold was chosen using a common criterion, i.e., the false positive rate is approximately 25%. AUCs obtained for the training set are reported in supplementary Table 3 along with optimal hyper-parameters used in each of the cross-validations.

**Table 1: Cross-validated goodness-of-fit metrics for SOZ determination.** Here we list the goodness-of-fit metrics (AUC, sensitivity, specificity, accuracy, precision, recall, and F1-score) obtained for the test dataset, for the different combinations of biomarkers and analytic techniques shown in Figure 4. The average values and standard deviations were computed using a) a tenfold stratified cross-validation and b) a leave-one-out cross-validation.

| Bio-marker | Method | AUC | Sensitivity (%) | Specificity (%) | Accuracy (%) | Precision (%) | Recall (%) | F1-score (%) |
|---|---|---|---|---|---|---|---|---|
| **10-fold CV** | | | | | | | | |
| All | SVM-LIN | 0.56(0.03) | 32.20(4.17) | 75.09(0.02) | 67.23(0.78) | 22.42(2.31) | 32.2(4.17) | 26.43(3.01) |
| All | SVM-RBF | 0.79(0.01) | 70.36(1.78) | 75.09(0.00) | 74.22(0.33) | 38.79(0.60) | 70.36(1.78) | 50.01(0.95) |
| HFO | SVM-RBF | 0.68(0.01) | 53.71(1.70) | 75.16(0.09) | 71.23(0.29) | 32.66(0.66) | 53.71(1.70) | 40.62(1.00) |
| IED | SVM-RBF | 0.68(0.01) | 55.11(2.53) | 75.07(0.03) | 71.41(0.45) | 33.14(1.01) | 55.11(2.53) | 41.39(1.50) |
| PAC | SVM-RBF | 0.73(0.01) | 60.63(2.74) | 75.09(0.02) | 72.44(0.51) | 35.31(1.03) | 60.63(2.74) | 44.63(1.57) |
| ALL | RATE | 0.58(0.01) | 35.91(2.03) | 75.09(0.02) | 67.91(0.37) | 24.43(1.03) | 35.91(2.03) | 29.07(1.40) |
| HFO | RATE | 0.56(0.01) | 32.39(2.20) | 76.31(0.62) | 68.26(0.77) | 23.48(1.46) | 32.39(2.20) | 27.22(1.74) |
| IED | RATE | 0.58(0.01) | 35.19(1.42) | 75.18(0.09) | 67.85(0.27) | 24.14(0.74) | 35.19(1.42) | 28.63(0.99) |
| PAC | RATE | 0.62(0.01) | 43.76(2.25) | 75.27(0.12) | 69.49(0.42) | 28.41(1.03) | 43.76(2.25) | 34.45(1.46) |
| **Leave one (subject) out CV** | | | | | | | | |
| ALL | SVM-RBF | 0.73(0.02) | 57.45(2.82) | 79.49(0.57) | 73.3(0.90) | 38.71(2.45) | 57.45(2.82) | 43.49(1.99) |
| HFO | SVM-RBF | 0.63(0.01) | 35.53(2.89) | 84.16(0.86) | 73.10(1.00) | 34.63(2.72) | 35.53(2.89) | 35.09(2.11) |
| IED | SVM-RBF | 0.60(0.01) | 33.50(2.22) | 80.77(0.66) | 69.47(0.89) | 30.55(2.42) | 33.50(2.22) | 29.16(1.55) |
| PAC | SVM-RBF | 0.69(0.01) | 47.70(2.81) | 81.03(0.63) | 72.63(0.93) | 36.23(2.39) | 47.70(2.81) | 39.06(1.94) |

*Combining Multiple Electrophysiological Biomarkers Improves the Localization Accuracy*

The classification framework depicted in Figure 3 was utilized with the features relevant to HFO, IED, and PAC biomarkers separately to reveal the predictive ability of individual biomarkers. Then the individual classification performances were compared against the performance obtained when all the biomarkers were used together. Figure 4d shows the ROC curves obtained for the different runs of the classification framework. While the PAC biomarker had the best predictive ability individually (AUC: 0.74 – 10-fold CV, 0.69 – leave-1-out CV), the classification obtained using all the biomarkers together performed better than any of the individual biomarkers, providing an AUC of 0.79 with the 10-fold CV approach and an AUC of 0.73 with the leave-1-out CV approach. These findings support the idea that the different interictal electrophysiological biomarkers used in this study possess complementary information that can be harnessed to achieve a superior performance in predicting the electrodes in an SOZ.

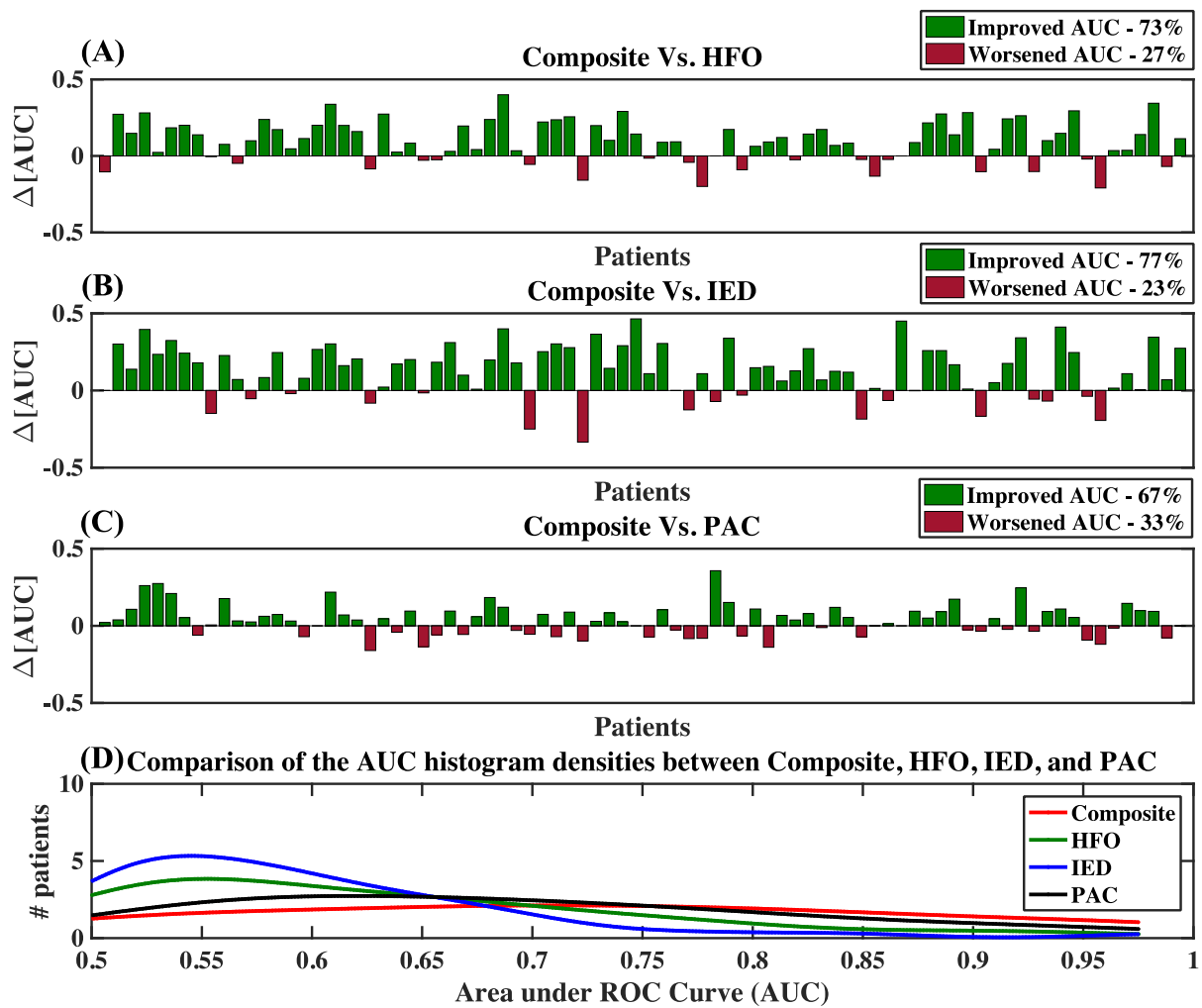

**Figure 5: Improvements in patient-specific SOZ classification achieved by means of combining multiple biomarkers as opposed to utilizing a single biomarker in the SVM framework.** (A), (B), and (C): Improvements obtained in AUCs for patient-specific SOZ classification when the combination of multiple biomarkers was utilized compared to when HFO, IED, or PAC, respectively, was utilized alone. (D) Histogram densities of the patient-specific AUCs for the prediction of SOZ electrodes using HFO, IED, PAC, and their composite.

To determine whether the combination of multiple biomarkers can reduce the inter-patient variability, we analyzed the improvements in SOZ electrode classification potential in each individual by calculating the AUC for each individual separately. We predicted the SOZ electrodes of each individual patient using HFO features, IED features, and PAC features separately and then their composite, within the SVM-based framework described previously using a leave-one-out cross-validation approach (see supplementary Table 2). Figures 5a–5c illustrate the respective improvements in the AUCs of individual patients attained when the combination of the three biomarkers was utilized instead of HFO, IED, or PAC by itself. Our analysis indicates that the AUCs of individual patients improved or remained unchanged for more than 65% of the patients when the composite was utilized compared to any individual biomarker. Then we plotted the histograms of the AUCs of individual patients for each biomarker and their composite separately, and approximated the densities by using kernel density estimation. Figure 5d shows that the histogram density of AUCs of individual patients becomes skewed towards higher AUC values when the composite is utilized to predict SOZ electrodes. This indicates that the utilization of multiple biomarkers with complementary information reduces the overall variability across patients in the ability to classify SOZ electrodes— variability that is apparent with any single biomarker.

*Recording Durations Between 90 and 120 Minutes May be Sufficient for Interictal SOZ Localization*

The ability to localize seizure-generating brain tissue is a cornerstone of clinical epileptology. We investigated how the duration of interictal iEEG recording impacted the the localization of the SOZ (as illustrated in Figure 6a). Here we applied our AI-based framework on a range of recording durations between 10 and 120 minutes. Figure 6 shows the mean ROC curves obtained using a tenfold cross-validation for different durations when an SVM classifier with an RBF kernel was utilized with all the biomarkers. To quantify the different runs, we plotted the AUC metric against the recording length used for SOZ electrode classification prediction. That is shown in Figure 6c, where the error bars indicate the standard deviations of the AUCs based on a tenfold cross-validation. Statistical significance tests using two-tailed paired t-tests indicate that the AUCs obtained using 90-, 100-, and 110-minute recordings are not statistically very different from the AUCs obtained using a 120-minute recording. This finding indicates that recording durations between 90 and 120 minutes may be sufficient for interictal SOZ identification with clinically relevant accuracies.

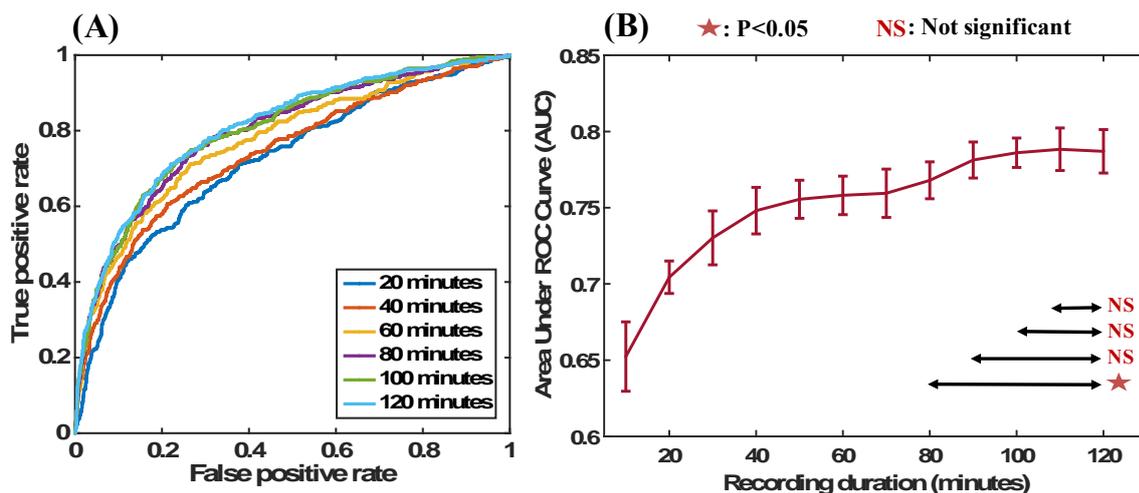

**Figure 6: Evaluation of the length of recordings and interictal SOZ localization.** (A) ROC curves obtained when shorter interictal segments of durations ranging from 20 to 120 minutes were utilized for analysis. (B) AUC values obtained with short interictal segments. Longer interictal segments result in better AUC values; however, the AUCs obtained using segments longer than 90 minutes are statistically indifferent (based on 2-sided paired t-tests between AUCs).

**Discussion**

*Main Contribution of the Study*

The current study describes a machine learning method for classification of SOZ and NSOZ electrodes using multiple electrophysiological biomarkers extracted from interictal iEEG data collected in a clinical setting. Our study, to our knowledge, is the first to utilize the complementary information provided by multiple electrophysiological biomarkers and their temporal characteristics as a way of reducing variability across patients to improve interictal SOZ localization. Using 2-hour wide-bandwidth intracranial EEG recordings of ~5000 electrodes from 82 patients, our study provides a large-scale evaluation of an artificial intelligence-based approach for interictal SOZ localization and shows that when used in concert with multiple interictal biomarkers, it can outperform single biomarker approaches. The ability to perform real-time feature processing and SOZ determination with clinically relevant accuracies, with a maximum monitoring duration of 2 hours, supports the feasibility of SOZ determination using interictal intracranial EEG data (see supplementary Table 2 for specific computation times for each task included in interictal SOZ classification). In addition to the above specific contributions, our study also exemplifies the role of artificial intelligence in augmenting clinical workflows.

*The Value of Combining Multiple Biomarkers*

Interictal SOZ localization techniques have been widely discussed, with the main focus being on the search for and validation of a single biomarker that can be used in all patients [Bragin 1999, Jacobs 2010, Jacobs 2008, Worrell 2011, Engel 2013]. Conventional methods have focused on HFO biomarker detection algorithms and on HFOs themselves [Worrell 2012, Burnos 2014, Balach 2014]. However, the generalizability of such single biomarkers has been insufficient for clinical practice [Nonoda 2016, Sinha 2017, Holler 2015, Cimbalnik 2017], primarily because of inter-patient variability, and it appears that one biomarker may not be sufficient to identify SOZs in all patients. While there have been multiple attempts to automate SOZ localization [Liu 2016, Graef 2013, Gritsch 2011], very little work has attempted to improve localization potential by means of combining multiple biomarkers. This exploratory study shows that combining multiple interictal electrophysiological biomarkers within a rigorous, supervised machine learning setting can be more accurate in performing interictal SOZ localization than can utilization of a single biomarker, essentially by reducing inter-patient variability. This is evident from our individual patient-based analysis (Figure 5), in which we show that SOZ electrode classification AUCs improve or remain unchanged for more than 65% of patients when the combination of the three biomarkers is utilized instead of single biomarkers. This finding indicates that combining multiple biomarkers reduces the variance in SOZ electrode classification and therefore achieves better generalizability than single-biomarker-based approaches. Figure 5 also shows that combining multiple biomarkers reduces the accuracy in SOZ electrode classification for some patients when compared with any of the individual biomarkers. The reason for this could be that there are more disagreements within the different biomarkers than agreements with respect to SOZ electrodes. The situations in which combining multiple biomarkers is detrimental can be explored and tested by comparing the individual predictions provided by the biomarkers.

*Exploiting the Temporal Variability in Epileptic Activity to Improve Classification Potential*

We showed in this study that utilizing a machine learning approach that uses local rates of biomarkers within 10-minute subintervals for classifying SOZ electrodes improves the AUC by 19% compared to traditional unsupervised approaches that primarily utilize the overall rates of biomarkers. We believe that the improvement is due to the inability of the overall-rate-based approaches to account for the temporal variations in epileptic activity. Recent studies have reported that the rate of epileptiform activity changes significantly between different behavioral states [Worrell 2008, Amiri 2016]. Since our study uses night segments with mixed behavioral states, the ability to differentiate SOZ electrodes from NSOZ electrodes using rates of epileptiform activity varies depending on the portion of the segment. Hence, looking at the overall rates of epileptiform activity might average out the variations between subintervals and hence degrade performance, and result in accuracy lower than the maximum attainable. We also show that a nonlinear classification technique provides significant improvements in AUC compared to a linear classification technique and that the performance provided by the latter is similar to that of commonly used unsupervised approaches. Although a linear classifier considers each subinterval for classification, it is impractical to assign a particular subinterval a higher weight because the exact subintervals in which the epileptiform activity is highly discriminative may not be the same across different patients. Therefore, when trained across multiple patients, the linear classifier perceives each subinterval as equally important and assigns all of them equal weights. As a result, its performance is similar to that of the approaches that use overall epileptiform activity rates to classify SOZ electrodes. On the other hand, an RBF kernel measures similarities (Euclidean distances) between a channel and selected SOZ and NSOZ channels (known as support vectors) with respect to their local rates of epileptiform activities. Therefore, regardless of the position of the subinterval that is highly discriminative between SOZ and NSOZ channels, that discriminative ability will be reflected in the overall distance between those channels.

That is pictorially illustrated in Figures 7a–7d using an example. Figure 7a shows that there is a large variation in the local rates of PAC (in 10-minute windows) across channels of a selected patient. Figure 7b shows the overall PAC rates for each channel, and shows that classifying SOZ electrodes simply by thresholding the overall rates results in poor sensitivity and specificity. Figure 7c shows that transforming the local rates to a kernel space using a linear kernel does not produce any major changes compared to the previous approach with regard to efficiency (in terms of sensitivity and specificity), whereas Figure 7d shows that transforming the local rates using an RBF kernel produces a more favorable transformation because it provides improved sensitivity and specificity. As shown in Figures 7c and 7d, kernel distances between feature vectors of individual channels and average feature vectors of the SOZ channels were calculated using linear and RBF kernels. This example elucidates the utility of a nonlinear machine learning classification approach and its relationship to the underlying physiology. It is also noteworthy that a nonlinear classification approach is suitable in this case due to the manner in which the features of channels were derived, and that linear classifiers may be sufficient for a different derivation of the features.

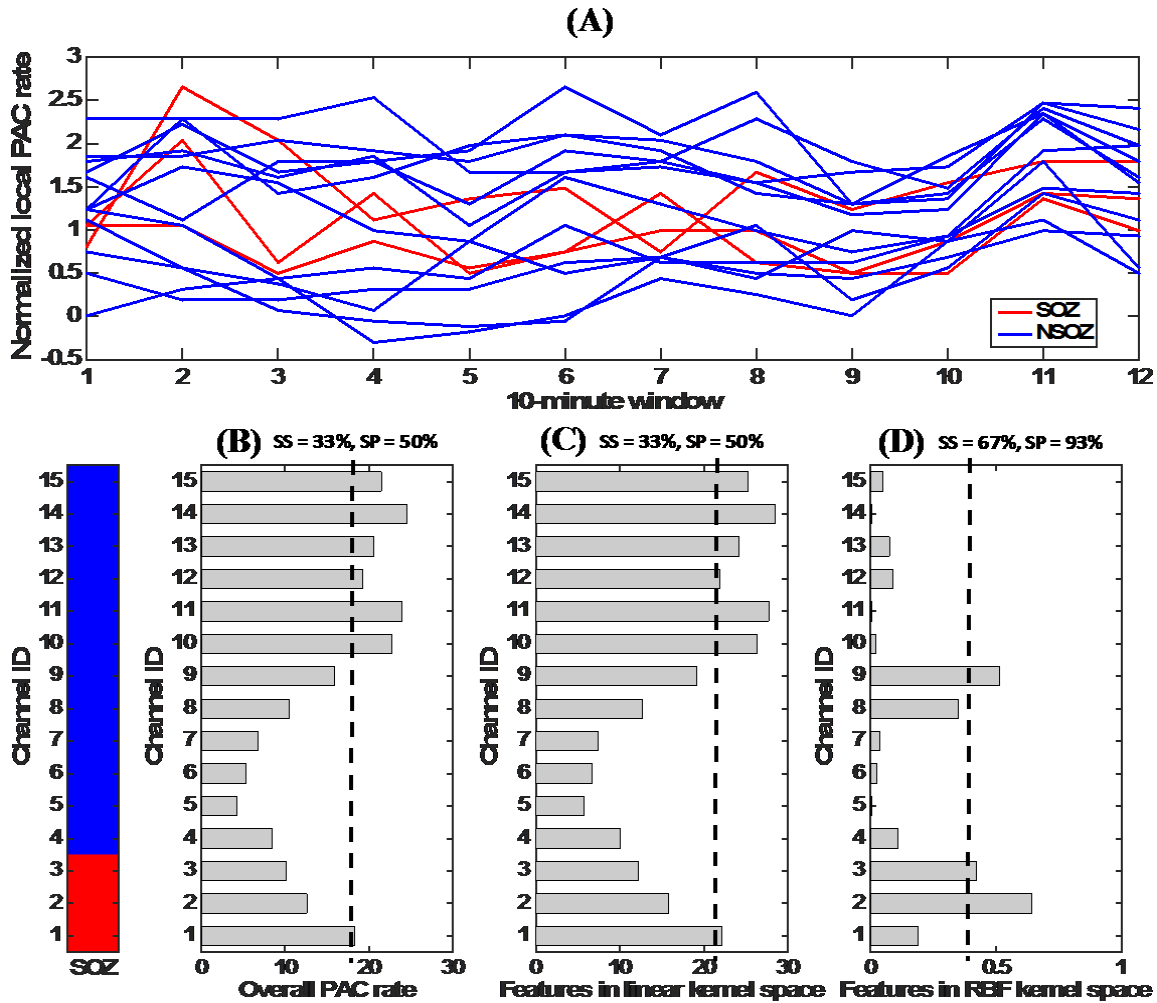

**Figure 7: Temporal variability in the rates of epileptiform activity (with respect to PAC) and its relation to nonlinear classification.** This figure illustrates that the RBF kernel is more specific in capturing the similarities between electrodes with regards to their epileptiform activity patterns. (A) Normalized local PAC rates in different 10-minute intervals of SOZ and NSOZ channels of a selected patient. (B) Overall PAC rates (obtained by summing local rates) for the channels and their poor ability to classify SOZ electrodes. (*SS* is sensitivity, *SP* is specificity, and a dashed line indicates application of a threshold). (C) Features after application of a transformation using the linear kernel. This transformation does not provide any notable improvement compared to the summation approach. (D) Features after application of a nonlinear transformation using the RBF kernel. It is evident that this transformation provides better discrimination between SOZ and NSOZ channels.

*Identifying SOZs During Long and Short Recordings of Night Segments*

In this study, we showed that localizing an SOZ using multiple features identified in 120 minutes of mixed behavioral state data provided accuracy similar to that obtained with shorter segments (see Figure 6). There is a clear relationship between the utility of this platform and the time of the recording used in analysis. Segments of 120 minutes were arbitrarily chosen to be the maximum amount of time a neurologist could use during an operating room recording to identify the SOZ. Interestingly, results as measured by the AUC did not significantly differ for recording durations between 90 and 120 minutes. It appears that, given satisfactory recording conditions, less than one hour may be all that is needed to achieve high SOZ identification accuracy with this platform. The relationship between the sleep-wake cycle and epileptiform discharges is well known [Sammaritano 1991, Staba 2002, Malow 1998]. Investigation of PAC, HFO, and IED using both short-term and long-term iEEG data shows an increment of the HFO rate with the non-rapid eye movement (NREM) sleep stage, especially in subjects with temporal lobe epilepsy [von Ellenrieder 2017, Amiri 2016, Staba 2004] [Engel 2009]. In addition, recent work has shown that the PAC localization potential increases in slow-wave sleep [Amiri 2016]. Our results suggest that obtaining sleep recordings of a sufficient duration in the clinical routine can be beneficial to quantitative SOZ localization.

*Contributions Towards Advancing the Current State of Clinical Decision-Making*

We demonstrated that our approach can identify electrodes in an SOZ with an accuracy of approximately 80% (AUC) on a large cohort of 82 patients, using interictal iEEG recordings of durations less than two hours. The idea of utilizing artificial intelligence to augment clinical workflows has become central in the era of "big data." Studies have shown the utility of deep-learning-based approaches in augmenting clinical diagnosis of skin cancer and diabetic retinopathy, primarily using imaging measurements [Esteva 2017, Gulshan 2016]. These studies have benefited considerably from the availability of a) huge image databases and b) substantially validated artificial neural-network-based classification models. Our study, on the other hand, embodies an alternative approach that utilizes feature engineering as a remedial option for the unavailability of large datasets at the scale of the currently available imaging datasets (with millions of medical images). We believe that the insights drawn from this study will be particularly useful for tasks that lack an abundance of labeled training samples, which inevitably is the case for most clinical problems.

*Study Limitations*

Our approach in its current implementation does not possess the ability to differentiate pathological and physiological electrophysiological events. For instance, HFOs are also associated with normal physiological function, and how to distinguish physiological HFOs from pathological HFOs is an active research area [Matsumoto 2013]. Recent studies have shown that utilizing only the pathological events results in increased sensitivity and specificity in determining SOZs interictally [Weiss 2016]. Hence, the ability to differentiate pathological events and then utilize them in determining SOZs may further improve our localization accuracy.

As shown recently [Amiri 2016], there is a clear connection between behavioral/sleep state and each of the biomarkers implemented in this study. We also showed that the temporal variations in epileptiform activity due to changes in behavioral states influences the machine learning paradigm utilized in this work. Therefore, accurate annotations of behavioral states considered together with interictal electrophysiologic biomarkers may further improve classification of epileptic and normal brain tissue. However, the patients in this cohort did not have scalp EEGs as required for accurate behavioral state classifications. New methods that can classify sleep stages based on intracranial recordings could prove useful for future analyses [Kremen & Duque 2016].

A successful clinical translation of this approach would depend on the accuracy of this approach in the data collected under operative settings. Prior studies evaluating IEDs [Wass 2001, Schwartz 1997] and HFOs [Zijlmans 2012, Wu 2010] for their potential to localize epileptic brain under intraoperative settings show promise. The PAC biomarker, however, despite being the best individual predictor in interictal settings, has not been evaluated under intraoperative settings and its specificity in such settings is still unclear. Hence, future studies evaluating multiple epilepsy patients are required to accurately determine the clinical utility of this approach.

Another limitation of our approach is that we take clinical SOZ as the gold standard in determining the accuracy of the interictal approach. This is a limitation because only a fraction of the patients going through epilepsy patients achieve complete seizure freedom (i.e., ILAE outcome 1). However, the goal of this study is to evaluate how accurately interictal biomarkers can localize the ictal recording SOZ localization. We make the assumption that any clues regarding SOZs that were not captured in ictal localization are less likely to be captured using interictal localization. Regardless, we observed that there is a good agreement between ictal and interictal SOZ localization approaches when the patients eventually had good outcomes (see supplementary Figure 1).

*Future Directions*

Advances in neuroimaging methods have advanced non-invasive seizure localization capabilities in epilepsy. Because of the holistic spatial view made possible by imaging techniques, they can provide independent information about potential regions of epileptogenic brain that could be used in concert with interictal electrophysiological biomarkers to further improve localization of pathological brain tissue. Alternatively, one can see the utility of combining pathological event classification, behavioral state classification, and multiple imaging modalities as another way of accounting for inter-patient variability, in addition to utilizing multiple interictal electrophysiological biomarkers. However, combining such different data types might require a more complex analytic technique in order to effectively capture the complementary information provided by each data type. We hypothesize that probabilistic graphical models provide an exceptional platform for handling such complexities, as shown in a recent work that combined spatial and temporal relationships in EEGs using a factor-graph-based model [Varatharajah 2017]. To that end, our future work will be directed towards harnessing the utilities of probabilistic graphical models in combining the forenamed multimodal data.

*Data and software availability*

The iEEG data and the software used in this study for AI-based interictal SOZ identification is available for download at ftp://msel.mayo.edu/EEG_Data/ and ftp://msel.mayo.edu/JNE_bundle.zip respectively.

**Conclusions**

Current methods for localizing seizure onset zones (SOZs) in drug-resistant epilepsy patients are lengthy, costly, and associated with patient discomfort and potential complications. Furthermore, among patients undergoing surgical resection for treatment of drug-resistant epilepsy, approximately 46% suffer seizure recurrence within 5 years. The current gold standard for localizing the tissue to be surgically resected is visual review of ictal (seizure) recordings. In this paper, we reported an artificial intelligence (AI) based approach for the prediction of SOZ electrodes using only non-seizure data. This technique was validated using interictal iEEG data clinically collected from 82 patients with drug-resistant epilepsy. The approach uses three pathological electrophysiological transients reported to be interictal (non-seizure) biomarkers of epileptic brain tissue: high-frequency oscillations (HFOs), interictal epileptiform discharges (IEDs), and phase-amplitude coupling (PAC). We utilize the complementary information provided by these 3 biomarkers and their temporal dynamics in a support vector machine classification (SVM) paradigm to reduce variability across patients and eventually to achieve an average area under ROC curve (AUC) measure of 0.73 in correctly classifying SOZ electrodes interictally. Furthermore, our results suggest that recording durations of approximately two hours are sufficient to localize the SOZ. Some potential applications of this technology are intra-operative electrode placement for prolonged monitoring, electrode placement for electrical stimulation devices [Fisher 2014] to treat seizures, and guiding of the margins of resection around epileptic structural lesions seen on MRI (magnetic resonance imaging). The potential to reduce the number of patients requiring long-term (multiple-day) iEEG monitoring is also interesting, but will require additional study. We postulate that in the future, combining pathological event classification, behavioral state classification, and multiple imaging modalities with interictal electrophysiological biomarkers could further improve interictal localization potential. We also believe that the insights drawn from this AI-based study will be particularly useful for clinical problems that require real-time data collection and decision making.